# On acceleration of Krylov-subspace-based Newton and Arnoldi iterations for incompressible CFD: replacing time steppers and generation of initial guess

Alexander Gelfgat

**Abstract** We propose two techniques aimed at improving the convergence rate of steady state and eigenvalue solvers preconditioned by the inverse Stokes operator and realized via time-stepping. First, we suggest a generalization of the Stokes operator so that the resulting preconditioner operator depends on several parameters and whose action preserves zero divergence and boundary conditions. The parameters can be tuned for each problem to speed up the convergence of a Krylov-subspace-based linear algebra solver. This operator can be inverted by the Uzawa-like algorithm, and does not need a time-stepping. Second, we propose to generate an initial guess of steady flow, leading eigenvalue and eigenvector using orthogonal projection on a divergence-free basis satisfying all boundary conditions. The approach, including the two proposed techniques, is illustrated on the solution of the linear stability problem for laterally heated square and cubic cavities.

## 1     Introduction

Stability of fluid flows, is one of the classical and oldest topics of theoretical fluid dynamics, has attracted much attention during the last decades. With the growth of computational power and fast development of numerical methods of linear algebra, it has become possible to study stability of numerically calculated flows. This requires the development of non-linear steady state solvers and the solution of eigenproblems for matrices of extremely large size. For a description of the topic and the problems being solved, the

School of Mechanical Engineering, Faculty of Engineering, Tel-Aviv University, Tel-Aviv, Israel
e-mail: gelfgat@tau.ac.il

reader is referred to review papers of Theofilis (2011), Dijkstra et al. (2013), and Juniper, Hanifi and Theofilis (2014). Existing computational methods and computer power can be used to study linear stability of two-dimensional flows relatively easily. Disturbances of these flows are usually assumed to be either two-dimensional or three-dimensional with prescribed spatial periodicity in one dimension, e.g., spanwise or circumferential. The family of such quasi-two-dimensional problems is called BiGlobal by Theofilis (2011) and Juniper, Hanifi and Theofilis (2014). Computational study of instability of fully three-dimensional flows without any preliminary assumptions about disturbances remains challenging for computational simulations. Difficulties are usually caused either by insufficiently powerful computer resources, or by an enormous slowdown of convergence of numerical methods successfully applied to the two-dimensional problems. Thus, development of robust numerical methods for three-dimensional stability problems is one of the most challenging problems of computational fluid dynamics.

One of the efficient and most popular approaches for application of Krylov-subspace linear algebra solvers to computation of incompressible flows and study of their stability was proposed by Tuckerman and Barkley (2000) and Tuckerman et al. (2000). Within this approach, assuming availability of an efficient time-dependent CFD code, Krylov vectors are generated via time stepping. Some necessary details regarding this are given below. This approach was successfully applied to a variety of two-dimensional problems, among which we cite only several recent ones (Boronska and Tuckerman, (2010a,b); Beaume, Bergeon and Knobloch (2011); Xin and Le Quéré (2012); Wang et al. (2014)) for example. Applying this technique to three-dimensional problems usually leads to a very slow convergence, so that the final result cannot be obtained in a reasonable time.

As is argued in Tuckerman and Barkley (2000) and Tuckerman et al. (2000), the time-stepping method for calculation of Krylov vectors can be interpreted as preconditioning by an inverse Stokes operator. The latter serves as a starting point for this study. Assuming a pressure/velocity coupled time integration we show how the Stokes preconditioning can be generalized. This allows for a faster convergence of the innermost iterative process, which produces Krylov vectors via iterative inversion of the preconditioned operator. Then, to reduce the number of outer iterations of either the Newton or Arnoldi solver, we propose to generate an initial guess which is close to the solution using projections on divergence-free bases. We argue also that having a good initial guess of the leading eigenvalue and eigenvector, the Arnoldi process can be replaced by a simpler and faster converging inverse iteration.

Application of the proposed approach is illustrated on the well-known problem of convection in square and cubic laterally heated cavities. For square cavities we reproduce the previously published results of Gelfgat (2007) for the critical Grashof number corresponding to the steady – oscillatory transition (Hopf bifurcation). Finally, we succeed in computing the critical Grashof numbers also for laterally heated three-dimensional cubic boxes with different thermal boundary conditions on horizontal and spanwise boundaries. The latter results are obtained by means of linear stability analysis for the first time.

In the following we consider a system consisting of momentum and continuity equations for an incompressible flow with velocity $\boldsymbol{u}$ and pressure $p$

$$\frac{\partial \boldsymbol{u}}{\partial t} = -\nabla p + \boldsymbol{L}(\boldsymbol{u}) + \boldsymbol{N}(\boldsymbol{u},\boldsymbol{u}) + \boldsymbol{f}, \quad div(\boldsymbol{u}) = 0 , \qquad (1,2)$$

where $\boldsymbol{L}$, $\boldsymbol{N}$ symbolize linear and bilinear operators, and $\boldsymbol{f}$ for density of bulk forces (see Dijkstra et al. (2013); Tuckerman and Barkley (2000); Tuckerman et al. (2000) for the details). It is assumed that the above equations are supplied with boundary conditions for all velocity components. We discuss two consecutive tasks: to calculate a steady state $(\boldsymbol{U}, P)$ of Eqs. (1) and (2) using the Newton iteration, and to calculate the leading eigenvalue of the problem linearized in the vicinity of the calculated steady state. The linearized time-dependent equations are

$$\frac{\partial \boldsymbol{u}}{\partial t} = -\nabla p + \boldsymbol{L}(\boldsymbol{u}) + \boldsymbol{N}_U(\boldsymbol{u}), \quad div(\boldsymbol{u}) = 0 . \qquad (3,4)$$

Clearly, the boundary conditions for Eqs. (3) and (4) are homogeneous, and are obtained after linearization (if needed) of the boundary conditions of Eqs. (1) and (2).

Considering the test problems, we show how a critical parameter corresponding to the instability threshold can be obtained by consecutive application of the two tasks described above.

## 1. Krylov-subspace-iteration-based Newton and Arnoldi methods via time steppers

In Krylov-subspace-based Newton and Arnoldi methods Krylov subspace iterations, BiCGstab(2) or GMRES (van der Vorst (2003)), usually are applied to compute the next Newton correction, or a next Krylov vector for the Arnoldi process (Edwards et al. (1994)). For the corresponding formulations and some details the reader is referred to Dijkstra et al. (2013), van der Vorst (2003), Edwards et al. (1994), and references therein. The Krylov basis vectors are defined by an initial vector $v^0$ and a matrix $A$ as $v^0, Av^0, A^2v^0, \ldots, A^nv^0, \ldots$. In Krylov subspace methods the solution of a linear algebraic equation system is approximated as a linear superposition of the $n$ Krylov vectors. Calculation of the Krylov basis is straightforward if the action of $A$ can be carried out on a vector, whether or not $A$ is represented as a matrix. When Newton's method or linear stability analysis are applied to incompressible flows, $A$ is the Jacobian matrix of eqs. (3,4). Its action on a vector $(\boldsymbol{u}, p)$ reduces to computation of the r.h.s of eq. (3), where pressure is evaluated to satisfy the constraint (4). Therefore, calculation of pressure for the next Krylov vector necessarily contains an implicit part, e.g., solution of the pressure Poisson equation (Edwards et al. (1994)). Also, for many incompressible flows, especially three-dimensional ones, the Jacobian matrix is ill-conditioned, which causes additional slowdown of convergence when its inverse is needed, e.g., for Newton and shift-and-invert Arnoldi iteration. Owing to the same reason conjugate-gradient type methods also experience slowdown of convergence.

The above computational problems can be partially overcome by the approach proposed by Tuckerman and Barkley (2000) and Tuckerman et al. (2000), based on the assumption that numerical time integration of Eqs. (1) and (2) is successfully realized. This approach is briefly described below.

For the following derivation, we follow Tuckerman and Barkley (2000) and assume that during time integration the pressure in Eq. (1) is obtained as a solution of the pressure Poisson equation with the right hand side bilinear with respect to the velocity $\boldsymbol{u}$. This allows us to incorporate the pressure gradient into the bilinear term $\boldsymbol{N}(\boldsymbol{u}, \boldsymbol{u}) = \boldsymbol{N}(\boldsymbol{u})\boldsymbol{u}$, so that a semi-implicit time integration scheme is defined as

$$\tfrac{1}{\delta t}\big(\boldsymbol{U}(t+\delta t)-\boldsymbol{U}(t)\big) = \boldsymbol{L}\mathbf{U}(t+\delta t) + \boldsymbol{N}\big(\mathbf{U}(t), \mathbf{U}(t)\big) + \boldsymbol{f}, \qquad (5)$$

from which the velocity at the next time step is evaluated as

$$\boldsymbol{U}(t+\delta t) = (I - \delta t \boldsymbol{L})^{-1}\big\{\boldsymbol{U}(t) + \delta t\big[\boldsymbol{N}\big(\boldsymbol{U}(t),\boldsymbol{U}(t)\big) + \boldsymbol{f}\big]\big\}. \qquad (6)$$

Using Eq. (6), the difference between two consecutive time steps can be expressed as

$$\boldsymbol{U}(t+\delta t) - \boldsymbol{U}(t) = (I/\delta t - \boldsymbol{L})^{-1}\big\{[\boldsymbol{N}\big(\boldsymbol{U}(t)\big)+\boldsymbol{L}]\boldsymbol{U}(t)+\boldsymbol{f}\big\}. \qquad (7)$$

Making the same assumptions about pressure, the next Newton correction $d\boldsymbol{u}$ is a solution of the following linear problem

$$(\boldsymbol{N}_U + \boldsymbol{L})d\boldsymbol{u} = [\boldsymbol{N}(\boldsymbol{U}) + \boldsymbol{L}]\boldsymbol{U} + \boldsymbol{f}, \qquad (8)$$

where $(\boldsymbol{N}_U + \boldsymbol{L})$ is the Jacobian matrix of the right hand side (r.h.s.) of Eq. (5). Using $(I/\delta t - \boldsymbol{L})^{-1}$ as a preconditioner, we observe that the r.h.s. of the preconditioned problem

$$(I/\delta t - \boldsymbol{L})^{-1}(\boldsymbol{N}_U + \boldsymbol{L})d\boldsymbol{u} = (I\delta t - \boldsymbol{L})^{-1}\{[\boldsymbol{N}(\boldsymbol{U}) + \boldsymbol{L}]\boldsymbol{U} + \boldsymbol{f}\} \qquad (9)$$

coincides with the r.h.s. of Eq. (7), so that it can be computed as a difference between two consecutive time steps.

Assuming that the linear problem (9) is solved using a Krylov subspace iteration method, each next Krylov vector $v^{n+1}$ is evaluated as

$$v^{n+1} = (I/\delta t - L)^{-1}(N_U + L)v^n. \qquad (10)$$

It is easy to see that the r.h.s. of Eq. (10) can be evaluated as a difference between two consecutive time steps of the linearized equations (2). Assuming $u(t) = v^n$,

$$v^{n+1} = u(t + \delta t) - u(t) = (I/\delta t - L)^{-1}\{[N_U(u(t)) + L]u(t)\}. \qquad (11)$$

The last equation can be applied also for calculation of the Krylov basis for the Arnoldi iteration if eigenvalues of the linearized equations (2) are needed. Clearly, to apply Eqs. (7) and (11) for Krylov-subspace based Newton or Arnoldi methods, one needs only velocity values computed by performing a time step.

Now, trying to generalize the above approach to other pressure/velocity segregated time steppers we note that the difference between two time steps in Eqs. (7) and (11) assume that the calculated velocity fields not only satisfy all the boundary conditions, but are also divergence-free. The latter means that the expression $(I/\delta t - L)^{-1}$ is not just a three-dimensional Helmholtz operator, since its action must result in a divergence-free velocity field satisfying all the boundary conditions. In the projection and fractional time step methods itimplicitly includes also parts for calculation of pressure and the correction step for velocity. Thus, for those semi-implicit time integration schemes, where only linear terms are calculated at the next time step, $(I/\delta t - L)^{-1}$ can be interpreted as the inverse Stokes operator (Tuckerman and Barkley (2000); Tuckerman et al. (2000)), whose definition will be given below.

Following Tuckerman et al. (2000), here are some additional comments. First, the inverse of the Stokes operator is considered here as a part of the time-stepping algorithm, so that at the end of each time step the boundary conditions are satisfied and the velocity is divergence-free. The latter is crucial for convergence of any Krylov subspace methods, since Krylov vectors must be also divergence-free and satisfy the boundary conditions, which are linear and homogeneous for the linearized problem. If the continuity equation is satisfied with insufficient accuracy, or boundary conditions are altered, the Krylov vectors will not belong to the correct linear space, and Krylov iterations will not converge. This can be observed by a straightforward numerical experiment. Second, the size of the time step $\delta t$ plays a role of the iteration parameter and must be chosen to maximize the convergence. Usually it is much larger than those used for the time integration. Recently, Beaume, Chini, Julien and Knobloch (2015) discussed how this parameter can be chosen basing on ratios of the time, diffusion and convection scales.

## 2 On possible extensions of the time stepper approach

Pressure/velocity segregated methods (e.g. fractional step, projection, influence matrix and others) are usually used for time integration of the incompressible momentum and continuity equations. For these approaches, the time step and the Stokes operator are an inherent part of discretization and/or computational algorithm, and cannot be altered. At the same time, pressure/velocity coupled methods, e.g., Edwards et al. (1994), Acharya et al. (2007), Feldman and Gelfgat (2009), Vitoshkin and Gelfgat (2013) and references therein, allow for some more freedom, which is discussed below. In the following we assume that all differential operators are approximated numerically by matrices, so that the numerical inverse of an operator is the inverse of the corresponding matrix. Using the notation of Vitoshkin and Gelfgat (2013), we represent the three-dimensional Stokes operator $S$ as a 4×4 operator matrix that acts on the vector of unknowns $(u, v, w, p)$ and produces the r.h.s. $(R_u, R_v, R_w, 0)$

$$S \begin{bmatrix} u \\ v \\ w \\ p \end{bmatrix} \equiv \begin{bmatrix} H_u & 0 & 0 & -\nabla_p^x \\ 0 & H_v & 0 & -\nabla_p^y \\ 0 & 0 & H_w & -\nabla_p^z \\ \nabla_u^x & \nabla_v^y & \nabla_w^z & 0 \end{bmatrix} \begin{bmatrix} u \\ v \\ w \\ p \end{bmatrix} = \begin{bmatrix} R_u \\ R_v \\ R_w \\ 0 \end{bmatrix}. \tag{12}$$

Here $\nabla^x$, $\nabla^y$ and $\nabla^z$ are the first derivatives in the *x*, *y* and *z* directions, $H = \Delta - I/\delta t$ are Helmholtz operators, $\Delta$ is the Laplacian operator, and $I$ is the identity operator. The lower indices show on which variable an operator acts. By assigning the lower indices, we emphasize that boundary conditions can be different for different velocity components, that different scalar variables can be assigned to different staggered grid nodes, and that numerical discretization of the same differential operators can be different for different scalar variables.

To discuss possible extensions of the time-stepping approach, we note that evaluation of the next Krylov vector (10) can be formally represented as

$$\begin{bmatrix} v^{n+1} \\ p^{n+1} \end{bmatrix} = M^{-1} \begin{bmatrix} (N_U + L)v^n \\ 0 \end{bmatrix} \tag{13}$$

where $M^{-1}$ is a preconditioning operator, and we arrive at the approach described above if $M = S$. An extension of the time-stepping approach would mean that $M \neq S$. Simultaneously, it means that the next Krylov vector cannot be obtained via time stepping, so that the numerical inverse of the operator $M$ is needed. Also, any preconditioner operator $M^{-1}$ must provide two essential properties of its action: the resulting velocity field must be divergence-free and must satisfy all the boundary conditions. Note that in the inverse Stokes operator (12), the divergence-free velocity results from its last row and column, while boundary conditions are incorporated into Helmholtz operators. Therefore, a straightforward extension will be a replacement of the Helmholtz operators by second order elliptic differential operators (all $\alpha$-s and $\beta$-s are positive)

$$Q_{(u,v,w)} = \alpha^x_{(u,v,w)} \frac{\partial^2}{\partial x^2} + \alpha^y_{(u,v,w)} \frac{\partial^2}{\partial y^2} + \alpha^z_{(u,v,w)} \frac{\partial^2}{\partial z^2} - \beta_{(u,v,w)} \ , \tag{14}$$

so that the preconditioner matrix becomes

$$M = \begin{bmatrix} Q_u & 0 & 0 & -\nabla^x_p \\ 0 & Q_u & 0 & -\nabla^y_p \\ 0 & 0 & Q_u & -\nabla^z_p \\ \nabla^x_u & \nabla^y_v & \nabla^z_w & 0 \end{bmatrix} . \tag{15}$$

The boundary conditions should be incorporated into operators $Q_{(u,v,w)}$, so that the inverse of $M$ yields a divergence-free field satisfying the boundary conditions, as does the inverse Stokes operator. The nine values of $\alpha$-s and three values of $\beta$-s must be optimized for the fastest convergence. Clearly, there are too many parameters to optimize just by a numerical experiment. However, some partial optimization is possible, and as is illustrated below, can be quite effective. It is stressed also that optimization of the preconditioner may speed up a Krylov-subspace iteration process, but it cannot alter, for example, the number of Newton iterations needed. It should be emphasized also that at this stage we depart from the time-stepping concept and compute the action of $(N_U + L)$ on a vector directly.

The inverse of matrix $M$ can be calculated by an Uzawa-like method proposed by Vitoshkin and Gelfgat (2013)[1]. First, the matrix $M$ is *LU* decomposed as

$$\begin{bmatrix} Q_u & 0 & 0 & -\nabla^x_p \\ 0 & Q_v & 0 & -\nabla^y_p \\ 0 & 0 & Q_w & -\nabla^z_p \\ \nabla^x_u & \nabla^y_v & \nabla^z_w & 0 \end{bmatrix} = \begin{bmatrix} I & 0 & 0 & 0 \\ 0 & I & 0 & 0 \\ 0 & 0 & I & 0 \\ \nabla^x_u Q_u^{-1} & \nabla^y_v Q_v^{-1} & \nabla^z_w Q_w^{-1} & I \end{bmatrix} \begin{bmatrix} Q_u & 0 & 0 & -\nabla^x_p \\ 0 & Q_v & 0 & -\nabla^y_p \\ 0 & 0 & Q_w & -\nabla^z_p \\ 0 & 0 & 0 & C \end{bmatrix}, \tag{16}$$

$$C = \nabla^x_u Q_u^{-1} \nabla^x_p + \nabla^y_v Q_v^{-1} \nabla^y_p + \nabla^z_w Q_w^{-1} \nabla^z_p \ . \tag{17}$$

Then the action of $M^{-1}$

---

[1] a similar result was proposed by Tau (1992)

$$\begin{bmatrix} u \\ v \\ w \\ p \end{bmatrix} = \boldsymbol{M}^{-1} \begin{bmatrix} R_u \\ R_v \\ R_w \\ 0 \end{bmatrix} \tag{18}$$

is calculated in three following steps

1. Solve $\hat{u} = Q_u^{-1} R_u$, $\hat{v} = Q_v^{-1} R_v$ and $\hat{w} = Q_w^{-1} R_v$ for $\hat{u}, \hat{v}$ and $\hat{w}$.
2. Solve $p = -C^{-1}\left(\nabla_x^u \hat{u} + \nabla_y^v \hat{v} + \nabla_z^v \hat{w}\right)$ for $p$.
3. Solve $u = \hat{u} + Q_u^{-1}\nabla_p^x p$, $v = \hat{v} + Q_v^{-1}\nabla_p^y p$, and $w = \hat{w} + Q_w^{-1}\nabla_p^z p$.

The elliptic operators $Q$ can be inverted either by multigrid or Krylov subspace iterations, or by the eigenvalue decomposition based TPF or TPT method, as proposed in Vitoshkin and Gelfgat (2013). The most time consuming step is the inverse of the matrix $C$. Here we note that action of this matrix on a scalar field $p$ can be represented as

$$Cp = div[\mathbf{Q}^{-1} grad p] = \nabla \cdot [\mathbf{Q}^{-1}\nabla p], \quad \mathbf{Q} = \begin{bmatrix} Q_u & 0 & 0 \\ 0 & Q_v & 0 \\ 0 & 0 & Q_w \end{bmatrix}. \tag{19}$$

Applying the inner product based on the volume integral ($V$ is the flow region and $A$ is its boundary)

$$\langle f, g \rangle = \int_V fg dV, \quad \langle \boldsymbol{u}, \boldsymbol{v} \rangle = \int_V \boldsymbol{u} \cdot \boldsymbol{v} dV, \tag{20}$$

we obtain

$$\langle Cp, p \rangle = \int_V p \nabla \cdot [\mathbf{Q}^{-1}\nabla p] dV = \int_V \nabla \cdot [p\mathbf{Q}^{-1}\nabla p] dV - \int_V \mathbf{Q}^{-1}\nabla p \cdot \nabla p dV. \tag{21}$$

The first integral in the above equality (21)

$$\int_V \nabla \cdot [p\mathbf{Q}^{-1}\nabla p] dV = \int_A [p\mathbf{Q}^{-1}\nabla p] dA = 0 \tag{22}$$

imposes velocity boundary conditions via action of $\mathbf{Q}^{-1}$, like in the step 3 of the above algorithm, and therefore this integral may vanish on the boundary $A$. This happens, for example, when all the boundary conditions are no-slip. Since the operator $\mathbf{Q}$, consisting of negative definite elliptic operators, is also negative definite, the second integral in Eq. (21) is negative, so that $\langle Cp, p \rangle > 0$, and the matrix $C$ is positive definite. The latter allows for application of Krylov subspace methods dedicated to positive definite matrices (van der Vorst, (2003)).

# 3     On implementing the Newton and Arnoldi iteration in "direct mode"

In this section we assume that all velocity boundary conditions are linear and uniform, e.g., no-slip conditions. Note that non-uniformity of a linear boundary condition can always be removed by a change of variables, so that only linearity is an essential requirement. Under the assumption made the solution belongs to a linear space $\mathcal{W}$ of divergence-free vectors satisfying all the boundary conditions. Following (Edwards, Tuckerman, Friesner, and Sorensen (1994)), we rewrite Eq. (13) as

$$v^{n+1} = \Pi(N_U + L)v^n. \tag{23}$$

We define $\Pi$ as a projection operator that projects a given 3D velocity vector on $\mathcal{W}$. If the projection operator is known, Eq. (23) yields a straightforward way for producing Krylov basis vectors for the inner iteration loop needed for both Newton and Arnoldi methods.

It is not clear, however, how to build this projection operator for a general numerical method. The Chorin projection and similar projectors used in the velocity/pressure segregated time steppers usually alter boundary conditions for tangent velocity. An example of such a projector was given in Edwards et al. (1994), where a pseudospectral method was applied. The projector is based on solution of the pressure equation derived by applying the divergence operator to the momentum equation (1) and including the no-slip boundary condition in the definition of the corresponding operator. A closer look shows that this approach is correct only when divergence and the Laplacian operator commute, which is true for spectral and pseudospectral methods with analytic evaluation of derivatives. When lower-order methods are implemented, e.g., finite differences, volumes or elements, the approximations of divergence and Laplacian operators do not commute near the boundaries, which makes the approach of Edwards et al. (1994) inapplicable.

An obvious version of the projector $\Pi$ is calculation of the orthogonal projection onto a set of basis functions $\{q_i\}_{i=1}^{\infty} \subset \mathcal{W}$. The functions $q_i$ must be divergence-free and satisfy all the homogeneous boundary conditions. As a rule, the basis functions are unknown, and must be defined for each problem separately. One possible way, based on linear superpositions of the Chebyshev polynomials was proposed in Gelfgat (2001, 2014), where all the definitions are given and technical details are described. This approach allows one to calculate coefficients $c_i$ of a truncated series

$$v^{n+1} = \Pi(N_U + L)v^n \approx \sum_{i=1}^{K} c_i q_i \tag{24}$$

by applying orthogonal (Galerkin) projections on the basis functions $q_i$. This process and the truncated sum (25) yields an approximation of the projection operator $\Pi$, if the truncation number $K$ is large enough. Here we note that the basis sets of Gelfgat (2001, 2014) are not orthogonal, which requires the computation and inversion of the Gram matrix. The latter task becomes too CPU-time consuming for large values of $K$, so that we propose here to apply these projections with relatively short truncations. This will produce a rough approximation of a true numerical solution within a shorter computational time. Then this rough approximation can be used as an initial guess either for computations with a larger $K$, or for the complete computation using the above preconditioner.

The whole computational process, which includes computation of (i) a steady flow and (ii) the corresponding leading eigenvalue, proceeds as follows. First the Newton method is applied in the projected mode (23) with the projections calculated as truncated sums (24) for gradually increasing truncation numbers, e.g., $10^3$, $20^3$, and $30^3$ for a 3D flow. The latter means approximation of the grid solution by 10 to 30 basis functions in each spatial direction. The result obtained for a smaller truncation number is used as an initial guess for the larger one. The approximate steady state obtained for the largest truncation is used as an initial guess for the Newton iteration preconditioned as in Eq. (18), which yields the converged numerical solution on a given grid. Since calculation of the leading eigenvalue at small truncations may be meaningless, only the largest truncation number is used to perform the Arnoldi iteration in the projected mode. Since this Arnoldi iteration is being run in the direct mode, it can be set to compute the eigenvalue with the largest real part (Sorensen (1992); Scott (1995)). The result yields an approximation of the leading eigenvalue and the eigenvector. The approximate leading eigenvalue defines the shift for the Arnoldi iteration in the shift-and-invert mode preconditioned by Eq. (18).

If the leading eigenvalue and eigenvector are known approximately, an alternative way to calculate the correct ones is by inverse iteration, which starts from the approximate eigenvector and is performed for the shifted Jacobian matrix. As above, the approximate eigenvalue defines the shift. Since inverse iteration converges to the eigenvector, with a good initial guess it can require lesser computational time than calculation of a representative Krylov basis needed for the Arnoldi approximation.

In many cases we need to calculate a critical parameter (e.g., Reynolds number $Re$) that corresponds to the leading eigenvalue having zero real part. In these cases, our outer computational loop solves equation $Real[\lambda(Re)] = 0$, for which we use the secant method. For each value of $Re$ we perform the above stages (i) and (ii), i.e. calculate the (i) corresponding steady flow, and (ii) the leading eigenvalue of the problem linearized near the computed steady state. For calculation of steady flow, we approximate the solution by truncated sums (24) when the current Reynolds number is noticeably different from the previous one, so that a good initial guess is unknown. When the secant method iterations approach the solution, the difference between the current and the previous Reynolds number becomes small, so that the previous solution is used to guess the next one. If the initial parameter ($Re$) is chosen close to from the critical one, the secant method converges in fewer than 10 iterations. Note that the secant method can be applied first to the projected equation, so that an approximation of the critical parameter is computed. Starting from this approximate parameter value, the full shift-and-invert mode for calculation of the leading eigenvalue is applied, and the secant method is restarted for computation of the final result.

## 4  Test problem

For the following numerical experiments we choose the same test problems as in Vitoshkin and Gelfgat (2013). We consider natural convection of an incompressible fluid in a 2D square or 3D cubic cavity, whose opposite sidewalls are kept at constant and different temperatures $T_{hot}$ and $T_{cold}$. The flow is described by Boussinesq equations that are rendered dimensionless taking the cube side length $H$ as a characteristic scale, and $H^2/\nu$, $\nu/H$, and $\rho \nu^2/H^2$ as scales of the time $t$, the velocity $\boldsymbol{v}$ and the pressure $p$, respectively. Here $\nu$ is the fluid kinematic viscosity and $\rho$ is the density. The temperature is rescaled to a dimensionless function using the relation $T \to (T - T_{cold})/(T_{hot} - T_{cold})$. Additionally, the dimensionless time, velocity and pressure are scaled, respectively by $Gr^{-1/2}$, $Gr^{1/2}$, and $Gr$, where $Gr = g\beta(T_{hot} - T_{cold})H^3/\nu^2$ is the Grashof number, $g$ is the gravitational acceleration and $\beta$ is the thermal expansion coefficient. The resulting system of momentum, energy and continuity equations reads

$$\frac{\partial T}{\partial t} + (\boldsymbol{v} \cdot \nabla)T = \frac{1}{PrGr^{1/2}} \Delta T \qquad (25)$$

$$\frac{\partial \boldsymbol{v}}{\partial t} + (\boldsymbol{v} \cdot \nabla)\boldsymbol{v} = -\nabla p + \frac{1}{Gr^{1/2}} \Delta \boldsymbol{v} + T\boldsymbol{e}_z \qquad (26)$$

$$\nabla \cdot \boldsymbol{v} = 0 \qquad (27)$$

Here $Pr = \nu/\alpha$ is the Prandtl number, and $\alpha$ is the thermal expansion coefficient. All the boundaries are assumed to be no-slip. Two vertical boundaries at $x = 0,1$ are kept isothermal, so that

$$T(x = 0, y, z) = 1, \quad T(x = 1, y, z) = 0 \qquad (28)$$

Other boundaries are assumed to be either perfectly thermally conducting or perfectly thermally insulated, which will be specified for each case separately. Further details can be found in Gelfgat (2009), Feldman (2011), and Vitoshkin and Gelfgat (2013), and references therein. In the following calculations the energy equation (25) was also preconditioned by an inverse elliptic operator

$$Q_T = \alpha_T^x \frac{\partial^2}{\partial x^2} + \alpha_T^y \frac{\partial^2}{\partial y^2} + \alpha_T^z \frac{\partial^2}{\partial z^2} - \beta_T \tag{29}$$

The equations were discretized by the finite volume method in space and three-level second order backward derivative in time. The corresponding Stokes-like preconditioner operators were inverted as described in the Section 3. All the details on the numerical approach used are given in Feldman and Gelfgat (2009), Feldman (2011), and Vitoshkin and Gelfgat (2013).

## 5 Some computational experiments

In all examples described below the calculations in full mode were done using the preconditioning matrix $M^{-1}$ that was inverted using the algorithm of Section 3. Calculations in the projected mode did not involve any preconditioning.

### 5.1 Two-dimensional problems

For a two-dimensional test problem we considered convection of air ($Pr = 0.71$) in a square laterally heated cavity with perfectly thermally insulated horizontal boundaries. All calculations in this section are carried out on a PC with an Intel® Core™ i7 dual processor.

First we examine convergence of the Newton method. The Newton correction $d\boldsymbol{u}$ in Eq. (8) was calculated using either BiCGstab($L$) (Slejipen and Fokkema (2003)) or FGMRES($m$) (van der Vorst (2003)). The inner iterations for solution of a linear equation system with the matrix $C$ (step 2 of the algorithm described in Section 3) were done using the ORTHOMIN(2) method (Zhang, Oyanagi, Sugihara (2004)). The convergence criteria was $|d\boldsymbol{u}/U| < 10^{-8}$ pointwise.

To examine convergence of the Newton method we calculate steady state flow at $Gr = 10^8$, using a calculated steady flow at $Gr = 5 \cdot 10^7$ as an initial guess on the staggered and stretched grid with $100 \times 100$ nodes. For the case considered and for all the preconditioners used, the BiCGstab($L$) method did not converge for $L$=2, 4, and 6. The FGMRES method always converged for $m = 300$.

The first calculation was carried out for the Stokes preconditioner $S^{-1}$, Eq. (12), applying $\delta t = 10$, which corresponds to the generalized preconditioner $M^{-1}$, Eq. (15), with all $\alpha$-s equal to unity and all $\beta$-s equal to 0.1. For this case, the Newton iterations converge in 650 sec. By varying $\alpha$-s and $\beta$-s, we found that the computational time can be reduced to 280 sec using all the $\beta$-s equal to $10^{-8}$ that corresponds to $\delta t = 10^8$. Further variation of these parameters causes of the total computational time to increase.

The problem with the above calculation is the FGMRES*(m)* method with a very large restart number, $m = 300$. Such a long restart may not be affordable due to memory restrictions for a 3D problem represented on a $100^3$ or finer grid. To use the BiCGstab(*L*) method, which requires less memory, we first apply the Newton iteration in the direct projected mode (see Section 4), calculating an approximate guess of the correct grid solution. For the $30^2$, $40^2$ and $50^2$ truncations it takes 95, 170, and 530 sec, respectively, using BiCGstab(4), while BiCGstab(2) does not converge. The steady state can be calculated applying BiCGstab(4) and using the approximate guess as the initial state. The calculation takes 480, 390, and 345 sec for initial states calculated by $30^2$, $40^2$ and $50^2$ truncation, respectively.

The next objective is convergence of the eigensolver and possible optimization of the preconditioner matrix parameters. For the following numerical experiments we calculated the leading eigenvalue $\lambda = (-0.03356, 0.67265)$, and its eigenvector at $Gr = 10^8$ by the Arnoldi and inverse iteration methods. For implementation of the Arnoldi method we used the ARPACK library with restarting after calculation of 20 Krylov vectors (Sorensen (1992)). The code for the inverse iteration method was written by the author. Using parameters optimized for the Newton method, i.e. all $\alpha$-s equal to unity and all $\beta$-s equal to $10^{-8}$ the calculations were completed in 6530 and 5400 sec for Arnoldi and inverse iteration methods, respectively. Changing all $\beta$-s to the value 0.1 reduces the times to, respectively, 5550 and 3780 sec. A series of further numerical experiments showed that these times can be decreased with the following choice of parameters: with $\alpha_u^x = \alpha_u^y = \alpha_v^x = \alpha_v^y = \alpha_T^x = 1$, $\alpha_T^y = 0.5, \beta_u = \beta_v = 3 \cdot 10^{-3}, \beta_T = 0.04$ the Arnoldi iteration converges in 3660 sec, and the inverse iteration converges in 2930 sec. We observe that the inverse iteration converges faster, which can be expected if the complex shift of the Jacobian matrix is close to the eigenvalue. In the above calculations the shift was $(0, 0.67)$.

The next two examples illustrate computation of the critical Grashof number. We use the preconditioner parameters found above without further optimization, and approximate projections on a subspace of divergence-free functions to obtain a good initial guess for the Newton method and inverse iteration. For the Arnoldi iteration in the direct projected mode, we use the EB13 routine of the HSL library, which allows also for Chebyshev acceleration of the starting vectors (Scott (1995)).

The first calculation of the critical Grashof number was performed for the cavity with perfectly thermally conducting horizontal walls, so that the temperature is prescribed there as

$$T(x, z = 0) = T(x, z = 1) = 1 - x \ . \tag{30}$$

This case is relatively easy since instability sets in before thin boundary layers are developed. The whole computation performed is detailed in Table 1. The steady state flow calculated at $Gr = 10^6$ was taken as the initial condition for following tests. To illustrate the proposed computation process, we calculate the steady state applying consecutive orthogonal projections on $10^2$, $20^2$, and $30^2$ divergence-free truncated bases, and use the last result as an initial guess to compute the steady state. In this way, the steady state at $Gr = 2.5 \cdot 10^6$ was calculated in 50.5 sec, which can be compared with 280 sec needed for the same calculation without the orthogonal projections approximation. After the steady solution is obtained, its eigenvalue is approximated by projecting the flow and all the Krylov vectors onto a $30^2$ basis. Here, under "number of main iterations" in Table 1, we report the number of restarts needed for convergence of the Arnoldi method that was restarted after computation of 100 Krylov vectors. The next calculation of steady flow and its leading eigenvalue is carried out for $Gr = 2.525 \cdot 10^6$. Since the Grashof number is altered only by 1% we do not need orthogonal projections to approximate the steady solution and use the formerly calculated flow as an initial guess (see Table 1). The two calculated eigenvalue approximations are used to linearly extrapolate the real part of the eigenvalue to zero, and to estimate the critical Grashof number value $Gr_{cr}$. Applying the secant method we obtain $Gr = 2.9018 \cdot 10^6$ for the next parameter value at which the steady state and leading eigenvalue should be computed. Repeating the whole sequence of calculations, we obtain the next approximation of $Gr_{cr}$ until arriving at $Gr_{cr} = 2.9387 \cdot 10^6$, for which approximation of the leading eigenvalue is (-7.952×10$^{-6}$, 1.587). The real part of the latter is considered as a numerical zero. At this stage we switch from calculation of approximate eigenvalues to computation of true ones. Applying the inverse iteration with the calculated complex shift and using the approximate eigenvector as the initial one, we obtain the leading eigenvalue $\lambda = (9.824 \cdot 10^{-7}, 1.587)$, whose real part appears to be even smaller than that of the approximate one. This means that approximation with $30^2$ basis functions is rather accurate for the case considered. The values obtained for $Gr_{cr}$ and $Im(\lambda)$ are in complete agreement with earlier results of Gelfgat (2007) obtained by a different computational approach. Note that most of computational time needed for calculations in the full mode is spent inversing the matrix $C$, which is the most CPU-time consuming part of the computational process and is analogous to calculation of pressure in time-dependent incompressible CFD.

Table 2 illustrates the whole computational process of calculation of the critical Grashof number in the case of perfectly thermally insulated horizontal boundaries of the square cavity

$$\frac{\partial T}{\partial z}(x, z = 0) = \frac{\partial T}{\partial z}(x, z = 1) = 0 \ . \tag{31}$$

The critical Grashof in this case is almost two orders of magnitude larger than the previous one (Gelfgat (2007)), so that convergence of all methods applied noticeably slows down. Consider, e.g., computation of steady flow at $Gr = 1.5 \cdot 10^8$, using the steady state at $Gr = 10^8$ as an initial guess. With orthogonal projection approximation using consequently $40^2$, $45^2$, and $50^2$ basis functions and the BiCGstab(4) method, the whole computation consumes 1520 sec. The same computation without orthogonal projections consumes 1920 sec using the BiCG*stab*(4) and only 330 sec using FGMRES(300). Starting from this value of the Grashof number it was impossible to obtain a computational process converging to $Gr_{cr}$. This happens because there are several eigenvalues with close real parts, so that the real part of the leading eigenvalue is no longer a smooth function of the Grashof number (see Table 2 and explanations below).

To obtain a converging computational process we had to start from $Gr = 2.1 \cdot 10^8$. To compute the initial steady state we have to perform parameter continuation with relatively small increments of $Gr$, so that first we compute the steady state at $Gr = 1.8 \cdot 10^8$, and only starting from it, the needed solution at $Gr = 2.1 \cdot 10^8$ is obtained. Using consequently $40^2$, $45^2$, and $50^2$ orthogonal projection approximations we arrive at the steady state in 2840 sec using BiCGstab(4) iterations for computation of Newton corrections, and alternatively in 2780 sec applying the FGMRES(300) method. Note that in this case the Newton iterations without orthogonal projections do not converge. Calculation of this steady state is followed by the calculation of $Gr_{cr}$. The whole process is reported in Table 2.

The computational process starts from calculation of the approximate leading eigenvalue at $Gr = 2.1 \cdot 10^8$ and at a 1% larger Grashof number $Gr = 2.121 \cdot 10^8$. The approximate leading eigenvalues are computed with $50^2$ truncation. In both cases we observe two leading eigenvalues with non-zero and zero imaginary parts respectively (Table 2). The real parts of these two eigenvalues approach zero at different rates, so that one, most unstable at lower $Gr$, is replaced by another one whose negative real part becomes larger, i.e., closer to zero. The estimation of $Gr_{cr}$ by the linear extrapolation cannot be correct here, therefore we repeat calculation of the steady state and its approximate eigenvalue at a close Grashof number, $Gr = 2.1527 \cdot 10^8$. The next linear extrapolation is rather good for the current leading eigenvalue with the zero imaginary part (Table 2), however, the next calculation at $Gr = 2.3198 \cdot 10^8$ reveals that a third eigenvalue has the largest real part. Its imaginary part is non-zero, but an order of magnitude smaller than that of the first one. This eigenvalue remains leading for the rest of the computational process. First we find the Grashof number at which the real part of the approximate leading eigenvalue vanishes, which is $Gr = 2.48418 \cdot 10^8$. Then, applying the inverse iteration, we find the value $Gr_{cr} = 2.26019 \cdot 10^8$, which is correct for the 100×100 grid considered, however it is not grid-converged yet (Gelfgat (2007)). Note that we did not need computations of steady states in the approximate projected mode, except the first one, since the difference between the current and the next Grashof number was smaller.

## 5.2      Three-dimensional problems

Three-dimensional calculations consume much longer CPU times, compared to the 2D ones, so that optimization of the preconditioner matrix parameters by a series of numerical experiments is not affordable. For the following calculations we extend optimal 2D parameters to the 3D case. Also, due to memory restrictions we cannot apply any of the GMRES algorithms with large restart numbers, and we use BiCGstab(4). As above, the Arnoldi method for projected solutions is used in the direct mode with restart after every 50 Krylov vectors and Chebyshev acceleration. As in the 2D case, the inner iterations for calculation of the matrix $C$ were done by the ORTHOMIN(2) method.

Table 3 presents an example of calculation of steady states and the critical Grashof number for convection in a cube with perfectly conducting horizontal and spanwise boundaries. With the boundary condition (28) on the lateral boundaries this reads

$$T(x, y = 0, z) = T(x, y = 1, z) = T(x, y, z = 0) = T(x, y, z = 1) = 1 - x. \quad (32)$$

Computation of the instability threshold starts from obtaining a steady flow at $Gr = 3.5 \cdot 10^6$ using the calculated steady state at $Gr = 2.8 \cdot 10^6$ as an initial guess. The BiCGstab($L$) iterations applied directly for calculation of the Newton corrections do not converge for $L$=2, 4, and 6. A convergent process was obtained by applying orthogonal projections on the bases of $10^3$, $20^3$ and $30^3$ functions (Table 3). We observe that most of the computational time was spent for calculation of an approximate solution on the $30^3$ basis, after which calculation of the steady state became faster and converged in a noticeably shorter time within a much smaller number of iterations.

The whole computational process reported in Table 3 is similar to those reported in Table 1 for the corresponding 2D case, but consumes significantly more CPU time on a computationally more powerful platform. It converges to $Gr_{cr} = 3.4136 \cdot 10^6$ with $\omega_{cr} = Im(\lambda) = 1.7651$. Results of similar calculations on the $100^3$ grid, where horizontal and spanwise boundaries are either thermally conducted or insulated, are summarized in Table 4. These results are in a good agreement with the results of time-dependent simulations (Janssen, Henkes and Hoogendoorn (1993); Labrosse, Tric, Khallouf, and Betrouni (1997); Soucasse, Riviére, Soufani, Xin, and Le Quéré (2014)). In all the three-dimensional calculations, we observe again that most of CPU time needed for calculation in the full mode is spent for inverting the matrix $C$. Finding a more efficient method for that is the main objective for further application of the proposed numerical technique to finer 3D grids.

## 6  Conclusions

We have proposed extensions of the time-stepping-based approach of Tuckerman and Barkley (2000) and Tuckerman et al. (2000) for application of the Krylov-subspace-based Newton and Arnoldi iterations to computation of steady incompressible flows and study of their stability. The approach of Tuckerman and Barkley (2000) and Tuckerman et al. (2000) is based on an existing code that performs numerical integration in time and is interpreted as preconditioning by the inverse Stokes operator. In the present study, we argue that convergence can be improved by modification of the Stokes operator. Similarly to the Stokes operator, this modification keeps Krylov vectors divergence-free and satisfying the boundary conditions, but also contains scalar parameters that can be tuned to speed up the convergence.

As an example, we propose to replace the Helmholtz operators, which are parts of the Stokes operator, by general elliptic operators leaving the boundary conditions untouched. Using two-dimensional convective flow as an example, we show that coefficients of the elliptic operators can be optimized so that the total computational time noticeably decreases. The resulting computational process is fully disconnected from any time-stepping algorithm. At the same time, the proposed preconditioner matrix can be inverted by methods applied for pressure/velocity coupled time integration, like the extended Uzawa method proposed by Vitoshkin and Gelfgat (2013), as well as direct and multigrid methods described in Feldman and Gelfgat (2009) and references therein.

We also argue that orthogonal projection of the whole problem on a truncated divergence-free basis yields divergence-free Krylov vectors, and allows one to perform Newton and Arnoldi iterations without additional implicit sub-steps. The projected solution is, in fact, an approximation of the true grid solution, and can be used as an initial guess for the complete computation. In particular, using projections allows for larger increments in parameter continuation calculations, and provides estimates of the leading eigenvalue and eigenvector needed to start either the Arnoldi or inverse iteration that are based on shifting and inverting the Jacobian matrix.

The proposed numerical approach is illustrated using the well-known problem of oscillatory convection of air in laterally heated square (2D) and cubic (3D) cavities. For 2D problems, we successfully recalculated previously published results of Gelfgat (2007). To the best of the author's knowledge, results for the 3D cavities are obtained by linear stability analysis for the first time and compare well with several results of direct numerical simulation.

**References:**

Acharya S, Baliga B.R, Karki K, Murthy JY, Prakash C, Vanka SP (2007) Pressure-based finite-volume methods in computational fluid dynamics. J Heat Transf., 129: 407–24.

Beaume C, Bergeon A, Knobloch E (2011) Homoclinic snaking of localized states in doubly diffusive convection, Phys. Fluids, 23: 094102.

Beaume C, Chini GP, Julien K, Knobloch E. (2015) Reduced description of exact coherent states in parallel shear flows, Phys. Rev. E, 91: 043010.

Borońska K Tuckerman LS (2010a) Extreme multiplicity in cylindrical Rayleigh-Bénard convection. I. Time dependence and oscillations. Phys. Rev. E, 81: 036320.

Borońska K, Tuckerman LS (2010b) Extreme multiplicity in cylindrical Rayleigh-Bénard convection. II. Bifurcation diagram and symmetry classification. Phys. Rev. E, 81: 036321.

Edwards WS, Tuckerman LS, Friesner RA, Sorensen DC (1994) Krylov methods for the incompressible Navier-Stokes equations, J. Comput. Phys., 110: 82-102.

Dijkstra HA, Wubs FW, Cliffe AK, Doedel E, Dragomirescu IF, Eckhardt B, Gelfgat AY, Hazel AL, Lucarini V, Salinger AG., Phipps ET, Sanchez-Umbria J, Schuttelaars H, Tuckerman LS, Thiele U (2013) Numerical bifurcation methods and their application to fluid dynamics: Analysis beyond simulation, Comm. Comput. Phys., 15: 1-45.

Feldman Y, Gelfgat AY (2009) On pressure-velocity coupled time-integration of incompressible Navier-Stokes equations using direct inversion of Stokes operator or accelerated multigrid technique, Computers & Structures, 87: 710-720.

Feldman Y (2011) Direct numerical simulation of transitions and supercritical regimes in confined three dimensional recirculating flows, PhD Thesis, Tel-Aviv University.

Gelfgat, AY (2001) Two- and three-dimensional instabilities of confined flows: numerical study by a global Galerkin method. Comput. Fluid Dyn. J., 9: 437-448.

Gelfgat, AY (2014) Visualization of three-dimensional incompressible flows by quasi-two-dimensional divergence-free projections, Computers & Fluids, 97: 143-155.

Gelfgat, AY (2007) Stability of convective flows in cavities: solution of benchmark problems by a low-order finite volume method, Int. J. Numer. Meths. Fluids, 53: 485-506.

Janssen RJA, Henkes RAWM, Hoogendoorn CJ (1993) Transition to time-periodicity of a natural convection flow in a 3D differentially heated cavity, Int. J. Heat Mass Transfer, 36: 2927-2940.

Juniper MP, Hanifi A, Theofilis V. (2014) Modal stability theory. Appl. Mech. Rev., 66: 024804.

Labrosse G, Tric E, Khallouf H, Betrouni M (1997) A direct (pseudo-spectral) solver of the 2D/3D Stokes problem: transition to unsteadiness of natural-convection flow in a differentially heated cubical cavity, Numer. Heat Transfer, Pt. B, 31: 261-276

Scott JA (1995) An Arnoldi code for computing selected eigenvalues of sparse real unsymmetric matrices, ACM Trans. Math. Software, 21: 432-475.

Slejipen LG, Fokkema DR (1993) BiCGstab(*L*) for linear equations involving unsymmetric matrices with complex spectrum. Electronic Transactions on Numerical Analysis, 1: 11-32.

Sorensen DC (1992) Implicit application of polynomial filters in a k-step Arnoldi method. SIAM J. Matrix Analysis and Applications, 13: 357-385.

Soucasse L, Riviére Ph, Soufani A, Xin S, Le Quéré P (2014) Transitional regimes of natural convection in a differentially heated cubical cavity under the effects of wall and molecular gas radiation, Phys. Fluids, 26: 024105.

Tau EY (1992) Numerical solution of the steady Stokes equations, J. Comput. Phys., 99: 190-195.

Theofilis V (2011) Global linear stability, Ann. Rev. Fluid Mech., 43: 319-3520.

Tuckerman LS, Barkley D (2000) Bifurcation analysis for time-steppers, In: Doedel K & Tuckerman L (eds) "Numerical Methods for Bifurcation Problems and Large-Scale Dynamical Systems", . IMA Volumes in Mathematics and Its Applications, 119, Springer, New York, p 453-466.

Tuckerman LS, Bertagnolio F, Daube O, Le Quéré P, Barkley D (2000) Stokes preconditioning for the inverse Arnoldi method, In . Henry D and Bergeon A (eds) "Continuation Methods for Fluid Dynamics" (Notes on Numerical Fluid Dynamics, 74), Vieweg, Göttingen, 2000, p 241-255.

Tuckerman L S (2015) Laplacian preconditioning for the inverse Arnoldi Method, Comm. Comput. Phys., 18: 1336-1351.

van der Vorst H (2003) Iterative Krylov Methods for Large Linear Systems. Cambridge Univ. Press, Cambridge.

Vitoshkin H, Gelfgat AY (2013) On Direct and Semi-Direct Inverse of Stokes, Helmholtz and Laplacian Operators in View of Time-Stepper-Based Newton and Arnoldi Solvers in Incompressible CFD, Commun. Comput. Phys., 14: 1103-1119.

Wang, B.-F, Wan, Z.-H, Ma, D.-J, Sun D.-J. (2014) Rayleigh-Bénard convection in a vertical annular container near the convection threshold. Phys. Rev. E., 89: 043014.

Xin S, Le Quéré P (2012) Stability of two-dimensional (2D) natural convection flows in air-filled differentially heated cavities: 2D/3D disturbances. *Fluid Dyn. Res.,* 44: 031419.

Zhang S-L, Oyanagi Y, Sugihara M. (2004) Necessary and sufficient conditions for the convergence of Orthomin(k) on singular and inconsistent linear systems. *Numerical Algorithms*, 36: 189-202

**Table 1:** Stability study of convection of air in a square laterally heated cavity with conducting horizontal walls. Calculations start from a steady state at $Gr=10^6$. 100×100 stretched grid. Preconditioner parameters: for Newton method: $\alpha_u^x = \alpha_u^y = \alpha_v^x = \alpha_v^y = \alpha_T^x = \alpha_T^y = 1, \beta_u = \beta_v = 10^{-8}$; For inverse iteration: $\alpha_u^x = \alpha_u^y = \alpha_v^x = \alpha_v^y = \alpha_T^x = 1, \alpha_T^y = 0.5, \beta_u = \beta_v = 3 \cdot 10^{-3}, \beta_T = 0.04$. Calculation on 2 Intel i7 2.93 GHz CPUs.

| Gr | stage | CPU time, sec | Number of main iterations | number of BiCGstab(4) iterations | time for inverse of C matrix | Calculated eigenvalue |
|---|---|---|---|---|---|---|
| $2.5 \times 10^6$ | Projection on 10×10 basis functions | 3.6 | 5 | 136 | | |
| | Projection on 20×20 basis functions | 7.1 | 3 | 237 | | |
| | Projection on 30×30 basis functions | 12.5 | 2 | 264 | | |
| | Full Newton iteration | 27.3 | 3 | 20 | 26.1 | |
| | Eigensolver in a direct projected mode | 32.7 | 41 | | | (-0.0236, 1.604) |
| $2.525 \times 10^6$ | Full Newton iteration | 36.2 | 2 | 30 | 30.9 | |
| | Eigensolver in a direct projected mode | 32 | 41 | | | (-0.0221, 1.603) |
| $2.9018 \times 10^6$ | Projection on 10×10 basis functions | 2.4 | 4 | 97 | | |
| | Projection on 20×20 basis functions | 8.9 | 4 | 279 | | |
| | Projection on 30×30 basis functions | 13.7 | 2 | 293 | | |
| | Full Newton iteration | 27.9 | 2 | 20 | 26.1 | |
| | Eigensolver in a direct projected mode | 32 | 41 | | | (-1.822×10$^{-3}$, 1.589) |
| $2.9355 \times 10^6$ | Full Newton iteration | 36.2 | 2 | 30 | 31.5 | |
| | Eigensolver in a direct projected mode | 32 | 41 | | | (-1.609×10$^{-4}$, 1.587) |
| $2.9387 \times 10^6$ | Full Newton iteration | 19.6 | 1 | 16 | 16.6 | |
| | Eigensolver in a direct projected mode | 32 | 41 | | | (-7.952×10$^{-6}$, 1.587) |
| | Inverse iteration in full mode | 575 | 4 | 194 | 534 | (9.824×10$^{-7}$, 1.587) |

**Table 2:** Stability study of convection of air in a square laterally heated cavity with insulated horizontal walls. Calculations start from a steady state at $Gr=2\cdot10^8$. 100×100 stretched grid. Preconditioner parameters: for Newton method: $\alpha_u^x = \alpha_u^y = \alpha_v^x = \alpha_v^y = \alpha_T^x = \alpha_T^y = 1, \beta_u = \beta_v = 10^{-8}$; For inverse iteration: $\alpha_u^x = \alpha_u^y = \alpha_v^x = \alpha_v^y = \alpha_T^x = 1, \ \alpha_T^y = 0.5, \beta_u = \beta_v = 3\cdot 10^{-3}, \beta_T = 0.04$. Calculation on 2 Intel i7 2.93 GHz CPUs.

| Gr | stage | CPU time, sec | Number of main iterations | number of BiCGstab(4) iterations | time for inverse of C matrix | Calculated eigenvalue |
|---|---|---|---|---|---|---|
| $2.1\times10^8$ | Projection on 40×40 basis functions | 1155 | 6 | 10630 | | |
| | Projection on 45×45 basis functions | 739 | 5 | 2464 | | |
| | Projection on 50×50 basis functions | 254 | 4 | 4859 | | |
| | Full Newton iteration | 688 | 3 | 571 | 661 | |
| | Eigensolver in a direct projected mode | 70 | 41 | | | (-0.02175, 3.4178) |
| $2.121\times10^8$ | Full Newton iteration | 1411 | 2 | 1221 | 1325 | |
| | Eigensolver in a direct projected mode | 263 | 167 | | | (-0.01309, 0) |
| $2.15275\times10^8$ | Full Newton iteration | 743 | 2 | 649 | 698 | |
| | Eigensolver in a direct projected mode | 602 | 397 | | | (-0.01306, 0) |
| $2.23198\times10^8$ | Full Newton iteration | 1167 | 3 | 998 | 1097 | |
| | Eigensolver in a direct projected mode | 495 | 326 | | | (-0.002529, 0.3158) |
| $2.25102\times10^8$ | Full Newton iteration | 686 | 2 | 605 | 643 | |
| | Eigensolver in a direct projected mode | 259 | 167 | | | ($0.4003\times10^{-3}$, 0.3154) |
| $2.248418\times10^8$ | Full Newton iteration | 383 | 1 | 338 | 359 | |
| | Eigensolver in a direct projected mode | 381 | 247 | | | ($0.5905\times10^{-5}$, 0.3154) |
| | Inverse iteration in full mode | 2263 | 5 | | 2129 | ($1.772\times10^{-2}$, 0.31488) |
| $2.250666\times10^8$ | Full Newton iteration | 288 | 1 | 260 | 269 | |
| | Inverse iteration in full mode | 3665 | 5 | | 3449 | ($1.4301\times10^{-2}$, 0.31483) |
| $2.260067\times10^8$ | Full Newton iteration | 1306 | 2 | 1161 | 1224 | |
| | Inverse iteration in full mode | 3643 | 7 | | 3134 | ($1.85608\times10^{-4}$, 0.31456) |
| $2.260190\times10^8$ | Full Newton iteration | 329 | 1 | 307 | 307 | |
| | Inverse iteration in full mode | 2715 | 5 | | 2567 | ($1.1227\times10^{-6}$, 0.31456) |

**Table 3:** Stability study of convection of air in a cubic laterally heated cavity with perfectly conducting horizontal and spanwise walls. Calculations start from a steady state at $Gr = 2.8 \cdot 10^6$. $100^3$ stretched grid. Preconditioner parameters for Newton method: $\alpha^{(x,y,z)}_{(u,v,w,T)} = 1, \beta_{(u,v,w,T)} = 10^{-8}$, and for inverse iteration: $\alpha^{(x,y,z)}_{(u,v,w)} = 1, \beta_{(u,v,w)} = 10^{-3}, \alpha^{(x,z)}_T = 1, \alpha^y_T = 0.5, \beta_T = 0.04$. Calculation on 32 AMD Abu Dhabi 2. 3 GHz CPUs.

| Gr | stage | CPU time, sec | Number of main iterations | number of BiCGstab(4) iterations | time for inverse of C matrix | Calculated eigenvalue |
|---|---|---|---|---|---|---|
| 3.5×10⁶ | Projection on 10³ basis functions | 1607 | 5 | 243 | | |
| | Projection on 20³ basis functions | 5178 | 4 | 590 | | |
| | Projection on 30³ basis functions | 37369 | 3 | 873 | | |
| | Full Newton iteration | 13135 | 5 | 120 | 11510 | |
| | Eigensolver in a direct projected mode | 272295 | 83 | | | (-0.7648×10⁻², 1.7609) |
| 3.5035×10⁶ | Full Newton iteration | 6370 | 2 | 27 | 5650 | |
| | Eigensolver in a direct projected mode | 301080 | 8287 | | | (-0.7407×10⁻², 1.7609) |
| 3.6110×10⁶ | Projection on 10³ basis functions | 6258 | 5 | 254 | | |
| | Projection on 20³ basis functions | 35623 | 4 | 605 | | |
| | Projection on 30³ basis functions | 269437 | 3 | 691 | | |
| | Full Newton iteration | 53290 | 5 | 153 | 49015 | |
| | Eigensolver in a direct projected mode | 289010 | 83 | | | (-0.2430×10⁻³, 1.7598) |
| 3.614655×10⁶ | Full Newton iteration | 8950 | 2 | 28 | | |
| | Eigensolver in a direct projected mode | 287713 | 83 | | | (-0.8109×10⁻⁵, 1.7598) |
| | Inverse iteration in full mode | 3614655 | 6 | 746 | | (+0.12275×10⁻², 1.7632) |
| 3.615017×10⁶ | Full Newton iteration | 4424 | 2 | 14 | 4030 | |
| | Inverse iteration in full mode | 32610 | 2 | 230 | 28790 | (+0.12296×10⁻², 1.7632) |
| 3.401940×10⁶ | Projection on 10³ basis functions | 5998 | 5 | 240 | | |
| | Projection on 20³ basis functions | 36595 | 4 | 571 | | |
| | Projection on 30³ basis functions | 334180 | 3 | 668 | | |
| | Full Newton iteration | 44539 | 5 | 138 | 425180 | |
| | Inverse iteration in full mode | 87204 | 5 | 515 | 77698 | (+0.74802×10⁻³, 1.7652) |
| 3.414158×10⁶ | Full Newton iteration | 9556 | 2 | 35 | 8566 | |
| | Inverse iteration in full mode | 55548 | 4 | 352 | 48942 | (+0.37743×10⁻⁴, 1.7651) |
| 3.413571×10⁶ | Full Newton iteration | 5490 | 2 | 17 | 5019 | |
| | Inverse iteration in full mode | 37467 | 3 | 216 | 33699 | (+0.22938×10⁻⁶, 1.7651) |

Table 4. Critical Grashof number and critical oscillation frequencies for buoyancy convection in a laterally heated cube calculated on $100^3$ stretched grid.

| Horizontal boundaries | Spanwise boundaries | $Gr_{cr}$ | $\omega_{cr} = Im(\lambda_{leading})$ |
|---|---|---|---|
| Conducting | Conducting | $3.4136 \cdot 10^6$ | 1.7642 |
| Conducting | Insulated | $3.3831 \cdot 10^6$ | 1.6489 |
| Insulated | Conducting | $1.2259 \cdot 10^8$ | 0.9579 |
| Insulated | Insulated | $4.2524 \cdot 10^7$ | 0.05395 |